\documentstyle[pre,aps,epsfig]{revtex} 

\newcommand{\app}{\rightarrow}

\newcommand{\df}{\Delta F}
\newcommand{\dfnp}{{\cal F}}

\newcommand{\ham}{{\cal H}}

\newcommand{\muz}{\hat{\mu}}

\newcommand{\sigz}{\hat{\sigma}}
\newcommand{\totd}{{\mathrm{d}}}
\newcommand{\xbf}{\textbf{x}}

\begin{document}
\title{Theory of Systematic Computational Error\\in Free Energy Differences}
\author{Daniel M. Zuckerman$^\ast$ and Thomas B. Woolf$^{\ast \dagger}$}
\address{$^\ast$Department of Physiology and  $^\dagger$Department of Biophysics,\\
Johns Hopkins University School of Medicine, Baltimore, MD 21205\\
\texttt{dmz@groucho.med.jhmi.edu, woolf@groucho.med.jhmi.edu}
}
\date{DRAFT!! \hspace{3em} \today \hspace{4em} DRAFT!!}
\maketitle

\begin{abstract}
Systematic inaccuracy is inherent in any computational estimate of a non-linear average, due to the availability of only a finite number of data values, $N$.
Free energy differences $\df$ between two states or systems are critically important examples of such averages in physical, chemical and biological settings.
Previous work has demonstrated, empirically, that the ``finite-sampling error'' can be very large --- many times $k_BT$ --- in $\df$ estimates for simple molecular systems.
Here, we present a theoretical description of the inaccuracy, including the exact solution of a sample problem, the precise asymptotic behavior in terms of $1/N$ for large $N$, the identification of universal law, and numerical illustrations.
The theory relies on corrections to the central and other limit theorems, and thus a role is played by stable (L\'{e}vy) probability distributions.
\end{abstract}

\vspace*{1cm}

\textbf{Introduction.} Free energy difference calculations have a tremendous range of applications in physical, chemical, and biological systems;
examples include computations relating magnetic phases, estimates of chemical potentials, and of binding affinities of ligands to proteins (e.g., \cite{Allen-Tildesley,Beveridge-1989,McCammon-1991,Kollman-1993,Frenkel-book,Landau-Binder}).
Since the work of Kirkwood \cite{Kirkwood-1935}, it has been appreciated that the free energy difference,
$\df \equiv \df_{0\app 1}$,
 of switching from a Hamiltonian $\ham_0$ to $\ham_1$
is given by a \emph{non-linear average},
\begin{equation}
\label{dfcomp}
\df = -k_B T \log{ \left [ \; \langle \, \exp{(-W_{0\app 1} / k_B T) } \, \rangle_0 \; \right ] } \, ,
\end{equation}
where 
$k_B T$ is the thermal unit of energy at temperature $T$ and
$W_{0\app 1}$ is the work required to switch the system from $\ham_0$ to $\ham_1$.
The angled brackets indicate an average over switches starting from configurations drawn from the equilibrium distribution governed by $\ham_0$.
In instantaneous switching the work is defined by 
$W_{0\app 1} = \ham_1(\xbf) - \ham_0(\xbf)$ for a start (and end) configuration $\xbf$;
however, gradual switches requiring a ``trajectory''-based work definition may also be used as was pointed out by Jarzynski \cite{Jarzynski-1997a,Jarzynski-1997b}.

\begin{figure}[h]

\begin{center}
\epsfig{file=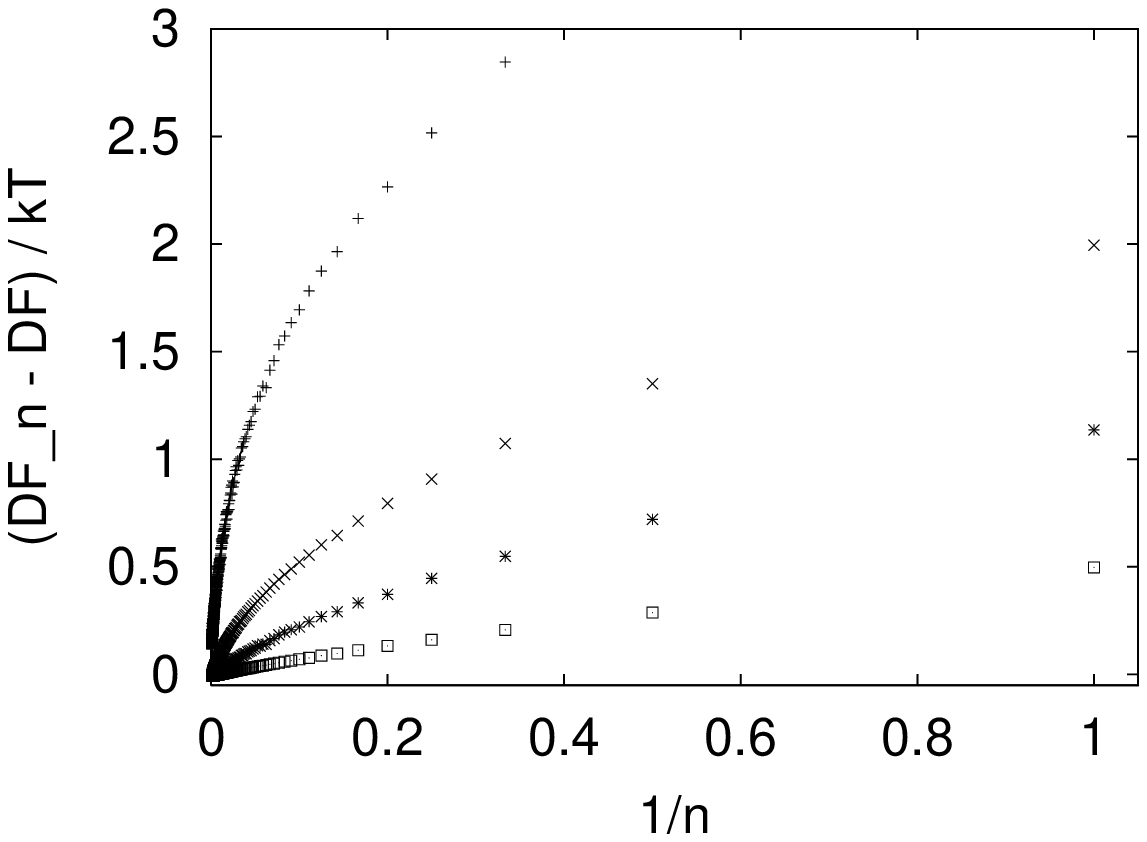,height=2.25in}
\epsfig{file=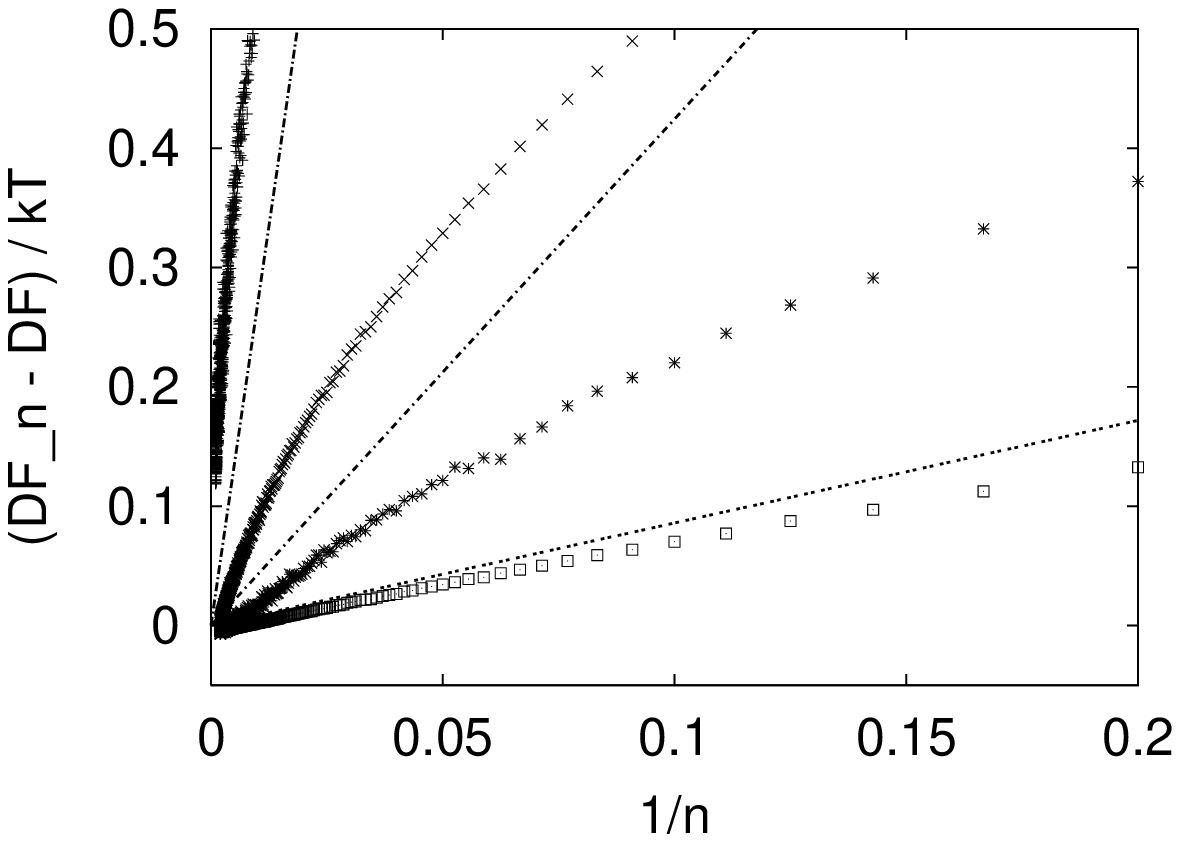,height=2.25in}
\end{center}

\caption{\label{fig:gauss}
Finite-sampling error for Gaussian-distributed work values.
The expected value of the dimensionless finite-sampling inaccuracy, $(\df_n - \df)/k_BT$ for $n$ data points is plotted as a function of $1/n$.
From top to bottom, the data sets represent numerical values of the error for Gaussian distributions of work values with standard deviations, $\sigma_w/ k_BT$ of 3, 2, 1.5, and 1.
The right panel also shows the exact, asymptotic linear behavior for the smallest widths.
}
\end{figure}

Whenever a convex, nonlinear average such as (\ref{dfcomp}) is estimated computationally, that result will \emph{always} be systematically biased \cite{Stone-1982} because one has only a finite amount of data --- say, $N$ work values. 
The bias results from incomplete sampling of the smallest (or most negative) $W_{0\app 1}$ values:
these values dominate the average (\ref{dfcomp}) and cannot be sampled perfectly for finite $N$, regardless of the $W_{0\app 1}$ distribution.
Thus, a running estimate of $\df$ will typically decline as data is gathered.
Such considerations led Wood \emph{et al.} \cite{Wood-1991} to consider the block-averaged $n$-data-point estimate of the free energy based on $N = m n$ total work values $\{W^{(k)} \}$, namely,
\begin{equation}
\label{dfncomp}
\df_n = \frac{1}{m} \sum_{j=1}^m 
          -k_B T \log{ \left[ \frac{1}{n} \sum_{k=(j-1)n+1}^{jn}
            \exp{ ( -W^{(k)} / k_B T ) } 
                       \right ] } \, .
\end{equation}
In the limit $m, N \app \infty$, $\df_n$ is mathematically well-defined and amenable to analysis;
it represents the expected value of a free energy estimate from $n$ data points --- that is, of
\begin{equation}
\label{dfninstance}
\dfnp_n = -k_BT \log{ \left[ \left. 
  \left( e^{-W_1/k_BT} + \cdots + e^{-W_n/k_BT} \right) \right/ n
  \right] } \, . 
\end{equation}
See Fig.\ \ref{fig:gauss}.
Wood \emph{et al.} estimated the lowest order correction to 
$\df \equiv \df_{\infty}$ as $\sigma_w^2 / 2 n k_B T$, where $\sigma_w^2$ is the variance in the distribution of work values, $W$ \cite{Wood-1991}.
Ferrenberg, Landau and Binder discussed analogous issues for the magnetic susceptibility \cite{Ferrenberg-1991}.

More recently, Zuckerman and Woolf \cite{Zuckerman-2001c} suggested a means by which a range of $\df_n$ values for $n < N$ could be used to extrapolate to the true, infinite-data answer, $\df$.
The authors also observed that, for large $m = N/n$, the free energy is bounded according to
\begin{equation}
\label{bound}
\df \leq \df_n \, , \hspace{0.5cm} \mbox{any } n \, .
\end{equation}
This inequality results from the convexity of the exponential function, as will be demonstrated explicitly in a fuller account of the theory.
Finally, Zuckerman and Woolf noted that the leading behavior of $\df_n$ appeared to be \emph{not always linear} in $1/n$ but, rather, seemed to behave as $(1/n)^{\tau_1}$ for $\tau_1 \leq 1$.

This Letter presents the theory --- apparently for the first time --- describing the finite-sampling inaccuracy for $\df$ estimates.
Previous work discussing $\df_n$ has been, primarily, empirical \cite{Wood-1991,Zuckerman-2001c}.
Our report includes 
(i) the formal analytic expression for the expected value of the error from $N$ work values, $\df_N - \df$;
(ii) an exact solution, for all $N$, of this expected value when the Boltzmann factor of the work value $z \equiv e^{-W/k_BT}$ follows a gamma distribution;
(iii) exact asymptotic expressions for $\df_n$ and the variance of $\dfnp$ as $n\app\infty$ for arbitrary $W$ distribtions, including non-analytic behavior in the case when the variance and higher moments of $z$ diverge; and 
(iv) discussion and numerical illustrations based on Gaussian distributions of $W$, plus corrections expected from skewed Gaussian distributions. 
The present discussion makes use of mathematical results regarding the convergence --- to ``stable'' limiting distributions \cite{Feller-1971,Zolotarev-1986,Uchaikin-Zolotarev}, also known as L\'{e}vy processes (e.g., \cite{Shlesinger-1995}) --- of the distributions of sums of variables.
The results are expected to have practical application in the extrapolation process outlined in \cite{Zuckerman-2001c}.

\textbf{Formal Development of $\df_n$.}
The derivation proceeds via continuum expressions simplified by the definitions
$w \equiv W/k_BT$, 
$f \equiv \df/k_BT$, and
$f_n \equiv \df_n/k_BT$.
 First, in terms of the probability density $\rho_w$ of work values, which is normalized by $\int \totd w \rho_w(w) = 1$,
the free energy is given by the continuum analog of (\ref{dfcomp}),
\begin{equation}
\label{dfint}
f = \df/k_BT  = - \log{ \! \left [ \int \!\! \totd w \, \rho_w(w) \, e^{-w}
                   \right ] } \, .
\end{equation}
The finite-data average free energy, following (\ref{dfncomp}) must apply the logarithm ``before'' the average of the $n$ Boltzmann factors, and one has
\begin{eqnarray}
\label{dfnint}
f_n & = & -\int \prod_{i=1}^n \left[ \totd w_i \, \rho_w(w_i) \right] \,
    \log{ \! \left [ \frac{1}{n} \sum_{i=1}^n e^{-w_i}
          \right ] } \, , 
\end{eqnarray}

Now, motivated by the central and related limit theorems \cite{Ash-1970,Feller-1971,Uchaikin-Zolotarev} for the sum of the $e^{-w}$ variables, we introduce a change of variables which will permit the development of a $1/n$ expansion for $f_n$.
In particular, we define
\begin{equation}
\label{ydef}
y = ( e^{-w_1} + \cdots + e^{-w_n} - n e^{-f}  ) \, / \, b_1 n^{1/\alpha} \, , 
\end{equation}
where $b_1$ is a constant and $\alpha \leq 2$ is an exponent characterizing the distribution of the variable $e^{-w}$. 
In fact, the requirement that $\df$ be finite in (\ref{dfint}) further implies $\alpha > 1$.
The finite-data free energy difference can now be written
\begin{equation}
\label{dfnofy}
f_n 
  = -\int_{-cn^a}^\infty \totd y \, \rho_n(y)  
      \log{ \! \left( e^{-f} + \frac{b_1}{n^a} y \right) } \, 
\end{equation}
where $c = \exp{(-f)} / b_1$,
$a \equiv (\alpha-1)/\alpha < 1/2$, and $\rho_n$ is the probability density of the variable $y$.
Note that $a$ is always positive because $\alpha>1$.


To continue, we must call upon some mathematical results regarding the approach, with increasing $n$, to general stable limit distributions (of which the Gaussian, for $\alpha=2$, is the best known \cite{Feller-1971,Uchaikin-Zolotarev}).
More precisely, the sum of \emph{any} set of random variables, suitably normalized as in (\ref{ydef}), has a distribution with zero mean which may be expressed as a stable distribution function multiplied by a large-$n$ asymptotic expansion \cite{Feller-1971,Christoph-Wolf}.

\textbf{Finite-Moments Case and An Exact Solution.}
To illustrate the case of a Gaussian limit ($\alpha=2$), assume the variable $e^{-w}$ possesses finite ``Boltzmann moments'' --- a mean $\muz=e^{-f}$, variance $\sigz^2$, and third moment $\muz_3$ --- not to be confused with the moments of the distribution of $w$.
The finite-$n$ corrections to the central limit theorem indicate that the variable $y = (\sum^n e^{-w_i} - n \muz)/\sqrt{n} \sigz$ [cf.\ (\ref{ydef})] is distributed according to \cite{Feller-1971}
\begin{equation}
\label{cltn}
\rho_n(y) = \rho_G(u;1)
  \left[ 1 + \nu_1(y) / \sqrt{n} + \nu_2(y) / n + \cdots
  \right], \,
\end{equation}
for large $n$, where the remaining terms are higher integer powers of $1/\sqrt{n}$ and the Gaussian density is
\begin{equation}
\label{guass}
\rho_G(y; \sigma) = \exp{(-y^2/2 \sigma^2)} / \sqrt{2 \pi} \sigma \, ,
\end{equation}
The $\nu_i$ depend on the original distribution of $e^{-w}$;
for instance, $\nu_1(y) = (\muz_3/6 \sigz^3) (y^3 - 3 y)$ \cite{Feller-1971}.  
Moreover, the $\nu$ functions are odd or even according to whether $i$ is odd or even, in this $\alpha=2$ case.

One arrives at the explicit form of the finite-data-corrected free energy for the case of finite $\sigz^2$ and $\muz_3$ by substituting (\ref{cltn}) into (\ref{dfnofy}), along with an expansion of the logarithm about $y=0$.
(More careful consideration of series convergence for large $y$ yields the same final result for $f_n$, as will be elucidated in future work.)
Because of the odd- and even-ness of the factors to be integrated, one finds an expansion consisting \emph{solely of integer powers of $1/n$}, namely,
\begin{equation}
\label{fnclt}
f_n = f + \varphi_1/n + \varphi_2/n^2 + \cdots \, ,
\end{equation}
with
$\varphi_1 = \sigz^2 / 2 \muz^2$ and
$\varphi_2 = -(4 \muz \muz_3 - 9 \sigz^4) / 12 \muz^4$ .
To compare this with the finding of Wood \emph{et al.} for $f_n - f$, one can consider a Gaussian distribution of $W = k_B T w$ with variance $\sigma_w^2$:
expanding the resulting Boltzmann moments of previous result for small $\sigma_w$ yields 
$\varphi_1  = k_BT [\exp{[(\sigma_w/k_BT)^2]} - 1] / 2 \approx \sigma_w^2 / 2 k_BT$, 
which yields precisely the first-order precdiction of Wood \emph{et al.} \cite{Wood-1991}.

Figure \ref{fig:gauss} illustrates the behavior of the finite-data free-energy for a Gaussian distribution of work values, based on numerical block averages (\ref{dfncomp}) and the asymptotic behavior given in (\ref{fnclt}).
Although the leading term in $f_n - f$ is linear in $1/n$, the leading coefficient is exponential in the \emph{square} of the distribution's width, while the next coefficient depends on the \emph{cube} of the width.
The asymptotic expressions (\ref{fnclt}) thus represent viable approximations only for a very small window about $1/n = 0$ for large widths.
Fig.\ \ref{fig:gauss} shows that such behavior is easily mistaken for non-analytic (e.g., power-law) behavior.

\begin{figure}[h]

\begin{center}
\epsfig{file=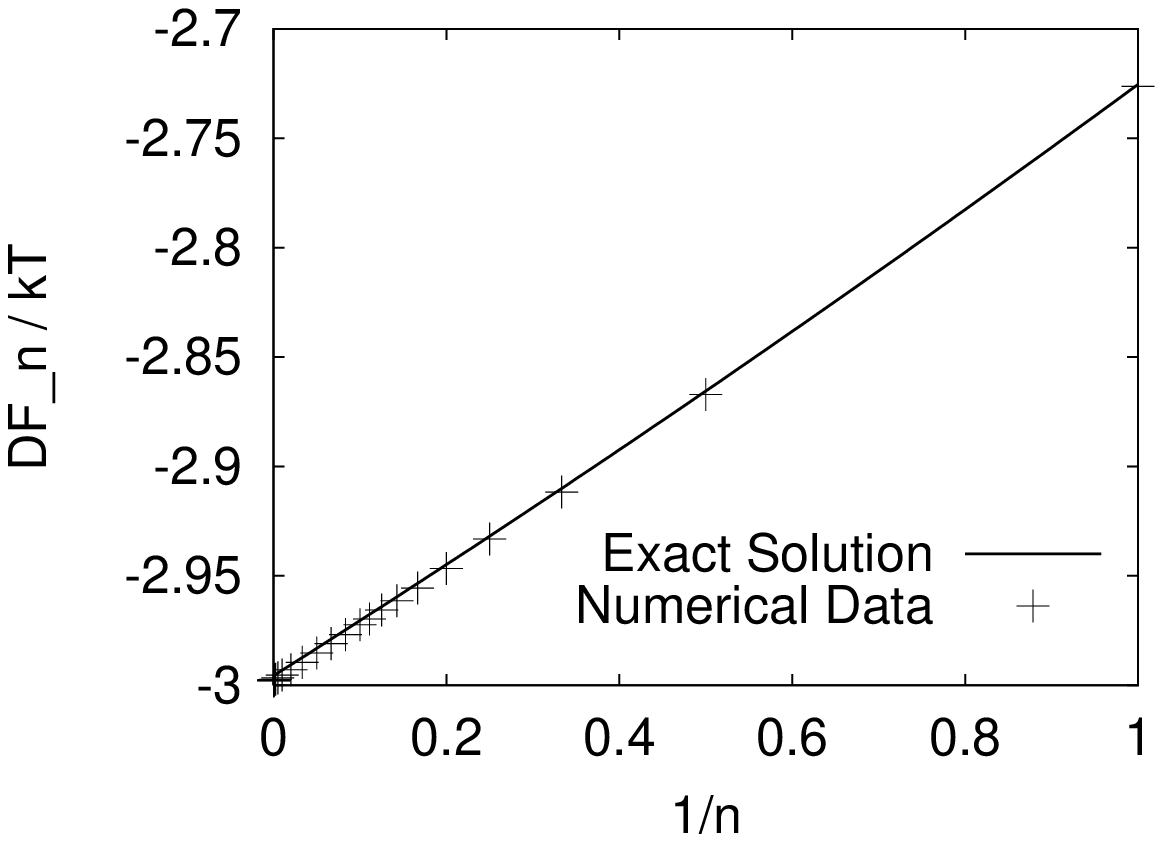,height=2.25in}
\epsfig{file=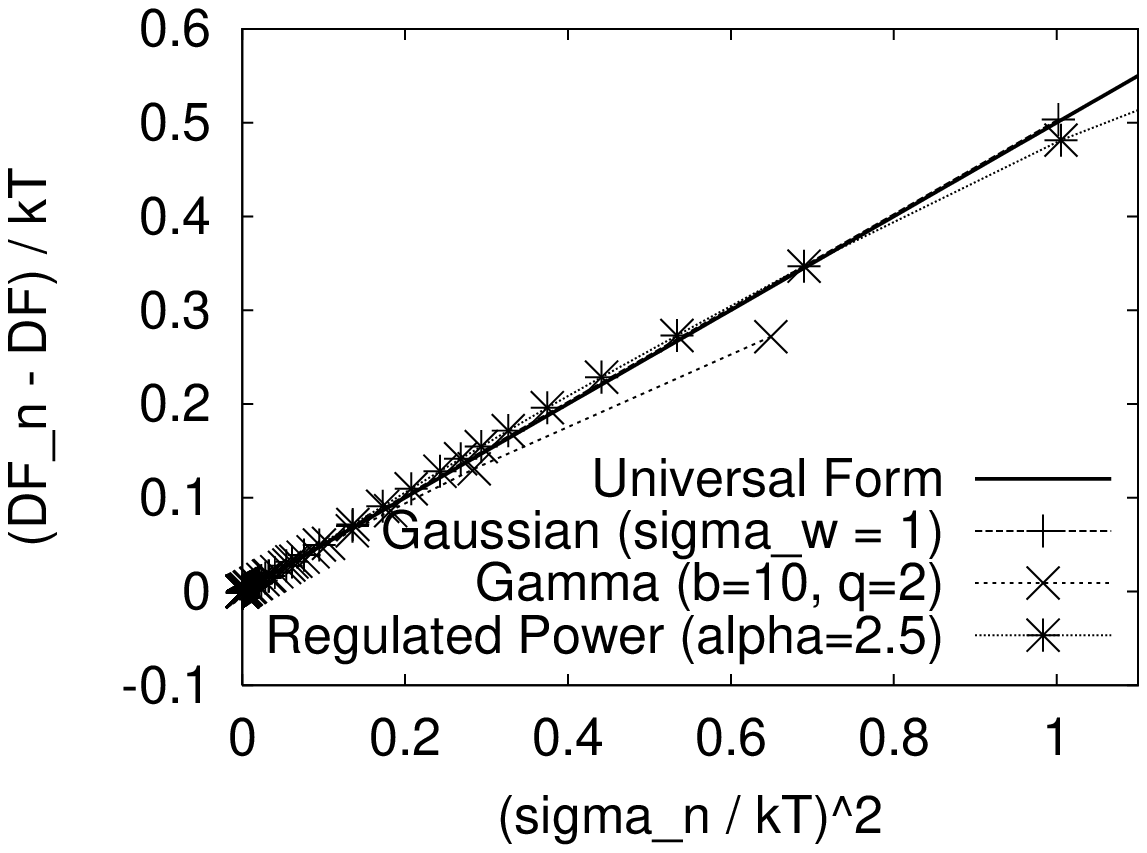,height=2.25in}
\end{center}

\caption{\label{fig:exact-uni}
An exact solution and a universal law.
The left panel illustrates the exact solution (\ref{dfn-gamma}) for the analytic form of $\df_n/k_BT$ when the work Boltzmann factor $e^{-W/k_BT}$ is distributed according to a gamma distribution (\ref{gamma-dist}).
The right plot illustrates the \emph{universal} asymptotic behavior of the finite-data free energy difference as a function of its fluctuations, $\sigma_n^2$;
see (\ref{finite-uni}) and text.
}
\end{figure}

An exactly solvable case occurs when the Boltzmann factor $e^{-w} \equiv z$ is distributed according to a gamma distribution, namely, 
\begin{equation}
\label{gamma-dist}
\rho_\Gamma(z;b,q) = (z/b)^{q-1} \exp{(-z/b)} \, / \, b \Gamma(q) \, .
\end{equation}
Because this density is ``infinitely divisible'' (see, e.g., \cite{Feller-1971}) the required sums in (\ref{dfninstance}) also follow gamma distributions, and after performing the integration described in (\ref{dfnofy}), one finds
\begin{equation}
\label{dfn-gamma}
f_n(n;b,q) = \log{(n/b)} - \psi(nq) \,
\end{equation}
where the digamma function is defined by
$\psi(x) = (\totd /\totd x) \Gamma(x)$.
The exact solution is illustrated in Fig.\ \ref{fig:exact-uni} for $b=10$, $q=2$.

When asymmetry is added to a Gaussian distribution via the first Edgeworth correction (see, e.g., \cite{Feller-1971}), one finds that the exponential dependence of the $\varphi_i$ on $\sigma_w$ is only corrected linearly by the now non-zero third moment of the $W$ distribution.

\textbf{Divergent Moments Case.}
When the variable $e^{-w} \equiv z$ in (\ref{ydef}) possesses a long-tailed distribution $\rho_z$, 
the limiting distribution is not a Gaussian and the results (\ref{cltn}) and (\ref{fnclt}) no longer hold.
In particular, if one of the tails of $\rho_z(z)$ decays as $z^{-(1+\alpha)}$ with $\alpha < 2$ (implying an infinite Boltzmann variance, $\sigz^2$), then the distribution of the variable $y$ in (\ref{ydef}) approaches a non-Gaussian stable law for large $n$ \cite{Uchaikin-Zolotarev}.   
Note that such power-law behavior in $z$ corresponds to \emph{simple exponential decay in the work distribution.}
Further, because the mean of $e^{-w}$ must be finite for $\df$ to exist [recall (\ref{dfint})], we also have $\alpha > 1$.
Unfortunately, no explicit forms for stable distributions are known in the range $1 < \alpha < 2$ \cite{Uchaikin-Zolotarev}.

A long-tailed $z$ distribution $\rho_z \equiv \rho_1$ also alters the \emph{form} of the asymptotic expansion of the sum-variable $y$ distribution and, hence, the expansion of $f_n$.
Instead of (\ref{cltn}), the distribution of the variable $y$ (\ref{ydef}) now takes the more complicated form \cite{Christoph-Wolf}
\begin{equation}
\label{stablen}
\rho_n(y) = \rho_\alpha(y) \left[
  1 + {\sum}^* \nu_{uv}(y) / n^{\theta(u,v)} 
  \right] \, , 
\end{equation}
where $\rho_\alpha$ is the appropriate stable probability density with exponent $\alpha$.
The functions $\nu_{uv}$, which are not available analytically, depend on the original distribution of $e^{-w}$ and partial derivatives of the stable distribution.
The exponents are given by
$\theta(u,v) = (u + \alpha v) / \alpha$,
and the summation ${\sum}^*$ includes $u\geq0$ and $v \geq - \lceil u/2 \rceil$, where $\lceil x \rceil$ denotes the integer part of $x$.
Note that we have omitted an asymmetry parameter, $\beta = 1$, of the stable laws \cite{Uchaikin-Zolotarev} which will be discussed in future work;
it does not, however, affect the form of the expansions.

Development of the expansion of $f_n$ for large $n$ in the case of diverging Boltzmann moments is more complicated, and will only be sketched here.
The basic strategy is to ensure that the coefficients of the powers of $1/n$ are all rendered in terms of convergent integrals, which requires both an expansion of the logarithm in (\ref{dfnofy}), as well as series \emph{and} asymptotic expansions of $\rho_\alpha$ in (\ref{stablen}) available from \cite{Feller-1971,Zolotarev-1986,Uchaikin-Zolotarev}.
The asymptotic result for $n \app \infty$ takes a reasonably simple form, namely,
\begin{equation}
\label{dfndiv}
f_n - f \approx \varphi_{\alpha-1} (1/n)^{(\alpha-1)} \, ,
\end{equation}
where 
$\varphi_{\alpha-1}(\alpha)$ depends on $\alpha$ and on the distribution $\rho_1$ in a complicated way; 
details will be presented in a future publication.

\textbf{Fluctuations and a Universal Law.}
The fluctuations in the finite-data free energy, $f_n = \df_n / k_BT$, as measured by the variance $\sigma_n$ of $\dfnp_n$ of (\ref{dfninstance}), are of considerable interest because of their potential to provide parameter-free extrapolative estimates of $f_\infty = \df/k_BT$ \cite{Zuckerman-2001c}.
Formally, the variance is given by
\begin{equation}
\label{sigmangen}
\left( \frac{\sigma_n}{k_BT} \right)^2 =
  \int_{-cn^a}^\infty \totd y \, \rho_n(y) \left[ \log{(1+y/cn^a)} \right]^2
  - ( f_n - f )^2 \, .
\end{equation}

Using techniques analogous to those sketched above yields asymptotic expansions for the fluctuations.
In the case of finite Boltzmann moments, one finds
\begin{equation}
\label{sigmanfinite}
\left( \sigma_n/k_BT \right)^2 \approx (\sigz / \muz )^2 / n 
  + O\left(n^{-2}\right) \, , 
\end{equation}
where it should be recalled that the unsubscripted moments refer to the density $\rho_z$.

Remarkably, comparison with $\varphi_1$ for (\ref{fnclt}) shows that
\begin{equation}
\label{finite-uni}
f_n - f = (\sigma_n / k_BT )^2 / 2 + O(n^{-2}) 
\end{equation}
exactly, as $n\app\infty$, and independent of any parameters of the distribution. 
This universal law, valid for the case when the second Boltzmann moment is finite, is illustrated in Fig.\ \ref{fig:exact-uni}.
The gamma distribution of Boltzmann factors was $\rho_\Gamma(z,10,2)$;
see (\ref{gamma-dist}).
The ``regulated power law'' distribution is defined by
$\rho_{rp}(z) = \alpha / (1+z)^{\alpha+1}$, and we set $\alpha=2.5$.

\textbf{Conclusions.}
Motivated by the need to understand the large-$N$ asymptotic behavior of free-energy-difference estimates based on a finite amount of data ($N$ work values), we have presented a general statistical theory which partially completes the task.
Two cases were formally identified, distinguished by whether the second moment of the distribution of \emph{Boltzmann factors} of the required work values is finite.
The asymptotic behavior was discussed for both cases, and 
--- for the finite-second-Boltzmann-moment case ---
an exact solution and a universal law were presented.

Much remains to be done, both in terms of theory and applications.
A question of particular practical interest is whether parameter-free extrapolation procedures can be devised, particularly in light of the sensitivity of the asymptotic behavior of $\df_n$ to the width of the distribution of work values.

\begin{acknowledgments}
The authors have benefitted greatly from discussions with Chris Jarzynski, Hirsh Nanda, Lawrence Pratt, and David Zuckerman.
We gratefully acknowledge funding provided by the NIH (under grant GM54782), the Bard Foundation, and the Department of Physiology.
D.M.Z. is the recipient of a National Research Service Award (GM20394) from the NIH.
\end{acknowledgments}


\end{document}